\documentstyle[12pt,epsf]{article}
\newcommand{\n}{\hspace*{-2.5mm}}
\newcommand{\li}{\mathop{{\mbox{Li}}_4}\nolimits}

\begin{document}
\title{
\vskip-3cm{\baselineskip14pt
\centerline{\normalsize\hfill FERMILAB--PUB--95/197--T}
\centerline{\normalsize\hfill MAD/PH/896}
\centerline{\normalsize\hfill MPI/PhT/95--65}
\centerline{\normalsize\hfill TTP95--28\footnote{The complete paper,
including figures, is also available via anonymous ftp at
ftp://ttpux2.physik.uni-karlsruhe.de/, or via www at
http://ttpux2.physik.uni-karlsruhe.de/cgi-bin/preprints/.}
}
\centerline{\normalsize\hfill hep-ph/9507382}
\centerline{\normalsize\hfill July 1995}
}
\vskip1.5cm
Virtual Top Effects on Low-Mass Higgs Interactions at Next-to-Leading Order in
QCD}
\author{{\sc Bernd A. Kniehl}\thanks{Permanent address:
Max-Planck-Institut f\"ur Physik, Werner-Heisenberg-Institut,
F\"ohringer Ring 6, 80805 Munich, Germany.}\\
{\normalsize Theoretical Physics Department, Fermi National Accelerator
Laboratory,}\\
{\normalsize P.O. Box 500, Batavia, IL 60510, USA}\\
{\normalsize and}\\
{\normalsize Department of Physics, University of Wisconsin,}\\
{\normalsize 1150 University Avenue, Madison, WI~53706, USA}\\ \\
{\sc Matthias Steinhauser}\\
{\normalsize Institut f\"ur Theoretische Teilchenphysik, Universit\"at
Karlsruhe,}\\
{\normalsize Kaiserstra\ss e 12, 76128 Karlsruhe, Germany}}
\date{}
\maketitle
\begin{abstract}
We present the next-to-leading-order QCD corrections of
${\cal O}(\alpha_s^2G_FM_t^2)$ to the low-$M_H$ effective $\ell^+\ell^-H$,
$ZZH$, and $W^+W^-H$ interaction Lagrangians in the high-$M_t$ limit.
In the on-shell scheme formulated with $G_F$, the
${\cal O}(\alpha_s^2G_FM_t^2)$ corrections support the
${\cal O}(\alpha_sG_FM_t^2)$ ones and further increase the screening of the
${\cal O}(G_FM_t^2)$ terms.
The coefficients of $(\alpha_s/\pi)^2$ range from $-6.847$ to $-16.201$, being
in line with the value $-14.594$ recently found for $\Delta\rho$.
All four QCD expansions converge considerably more rapidly, if they are written
with $\mu_t=m_t(\mu_t)$, where $m_t(\mu)$ is the $\overline{\mbox{MS}}$ mass,
rather than the pole mass, $M_t$.

\end{abstract}

\newpage

Now that the existence of the top quark has been established \cite{abe}, the
Higgs boson is the last missing link in the Standard Model (SM).
The discovery of this particle and the study of its properties are among the
most urgent goals of present and future high-energy colliding-beam experiments.
The Higgs boson is currently being searched for with the CERN Large
Electron-Positron Collider (LEP1) and the SLAC Linear Collider (SLC) via
Bjorken's process \cite{bjo}, $e^+e^-\to Z\to f\bar fH$.
At the present time, the failure of this search allows one to rule out the
mass range $M_H\le64.3$~GeV at the 95\% confidence level \cite{jan}.
The hunt for the Higgs boson will be continued with LEP2 via the
Higgs-strahlung mechanism \cite{ell}, $e^+e^-\to ZH\to f\bar fH$.
In next-generation $e^+e^-$ linear supercolliders (NLC), also
$e^+e^-\to\bar\nu_e\nu_eH$ via $W^+W^-$ fusion and, to a lesser extent,
$e^+e^-\to e^+e^-H$ via $ZZ$ fusion will provide copious sources of
Higgs bosons.

Once a novel scalar particle is discovered, it will be crucial to decide if it
is the very Higgs boson of the SM or if it lives in some more extended Higgs
sector.
For that purpose, precise knowledge of the SM predictions will be mandatory,
{\it i.e.}, quantum corrections must be taken into account.
The status of the radiative corrections to the production and decay processes
of the SM Higgs boson has recently been reviewed \cite{pr}.
Since the top quark is by far the heaviest known elementary particle,
with a pole mass of $M_t=(180\pm12)$~GeV \cite{abe}, the leading high-$M_t$
terms, of ${\cal O}(G_FM_t^2)$, are particularly important, and it is desirable
to gain control over their QCD corrections.
During the last year, a number of papers have appeared in which the two-loop
${\cal O}(\alpha_sG_FM_t^2)$ corrections to various Higgs-boson production and
decay processes are presented.
The list of these processes includes
$H\to f\bar f$, with $f\ne b$ \cite{hll} and $f=b$ \cite{ks1,kwi},
$Z\to f\bar fH$ and $e^+e^-\to ZH$ \cite{ks2},
$e^+e^-\to\bar\nu_e\nu_eH$ via $W^+W^-$ fusion \cite{ks3},
$gg\to H$ \cite{ks3,gam},
and more \cite{ks3}.
In this paper, we shall proceed one step beyond and tackle with three-loop
${\cal O}(\alpha_s^2G_FM_t^2)$ corrections.
To simplify matters, we shall work in the limit $M_H\ll M_t$ and concentrate on
reactions with colourless external legs.
Such reactions typically involve the $\ell^+\ell^-H$, $W^+W^-H$, and $ZZH$
couplings together with the gauge couplings of the $W$ and $Z$ bosons to the
leptons.
Our primary task is thus to find the next-to-leading QCD corrections to the
low-$M_H$ effective $\ell^+\ell^-H$, $W^+W^-H$, and $ZZH$ interaction
Lagrangians.

Recently, the ${\cal O}(\alpha_s^2G_FM_t^2)$ correction to $\Delta\rho$ has
been calculated and found to be quite sizeable \cite{avd},
being right at the edge of affecting ongoing precision tests of the standard
electroweak theory.
For $N_c=3$ and $n_f=6$, the QCD expansion of $\Delta\rho$ reads \cite{avd,djo}
\begin{equation}
\Delta\rho=3X_t\left[1-2.859\,912\,a\left(1+{7\over4}aL\right)
-14.594\,028\,a^2\right],
\end{equation}
where $a=\alpha_s(\mu)/\pi$, $X_t=(G_FM_t^2/8\pi^2\sqrt2)$,
$L=\ln(\mu^2/M_t^2)$, $G_F$ is Fermi's constant, and $\mu$ is the QCD
renormalization scale.
It is of great theoretical interest to find out whether the occurrence of
significant ${\cal O}(\alpha_s^2G_FM_t^2)$ corrections is specific to
$\Delta\rho$ or whether this is a common feature among the electroweak
parameters with a quadratic $M_t$ dependence at one loop.
In the latter case, there must be some underlying principle which is
responsible for this phenomenon.
Our analysis will put us into a position where we can investigate this issue
for four independent quantities.

We shall work in the electroweak on-shell renormalization scheme, with $G_F$
as a basic parameter \cite{sir}.
We shall take the colour gauge group to be SU(3), so that $N_c=C_A=3$,
$C_F=4/3$, and $T_F=1/2$.
We shall explicitly include five massless quark flavours plus the massive top
quark in our calculation, so that we have $n_f=6$ active quark flavours
altogether.
We shall evaluate the strong coupling constant, $\alpha_s(\mu)$, at
next-to-leading order in the modified minimal-subtraction
($\overline{\mbox{MS}}$) scheme \cite{msb}.
The $W$-, $Z$-, and Higgs-boson self-energies $\Pi_{WW}(q^2)$, $\Pi_{ZZ}(q^2)$,
and $\Pi_{HH}(q^2)$ will be the basic ingredients of our analysis.
In the case of $\Pi_{HH}(q^2)$, we shall actually need the first derivative
$\Pi_{HH}^\prime(q^2)$ for the Higgs-boson wave-function renormalization.
Since we wish to extract the leading high-$M_t$ terms, we may put $q^2=0$.
While the ${\cal O}(\alpha_s^2G_FM_t^2)$ results for $\Pi_{WW}(0)$ and
$\Pi_{ZZ}(0)$ are now well established \cite{avd}, $\Pi_{HH}^\prime(0)$
requires a separate analysis, which we shall carry out here.
Our calculation will proceed along the lines of Ref.~\cite{avd}.
We shall present our main results in this letter.
The technical details and a variety of applications will be reported elsewhere
\cite{long}.

\begin{figure}[ht]
 \begin{center}
 \begin{tabular}{ccc}
   \epsfxsize=5.0cm
   \leavevmode
   \epsffile[130 260 470 530]{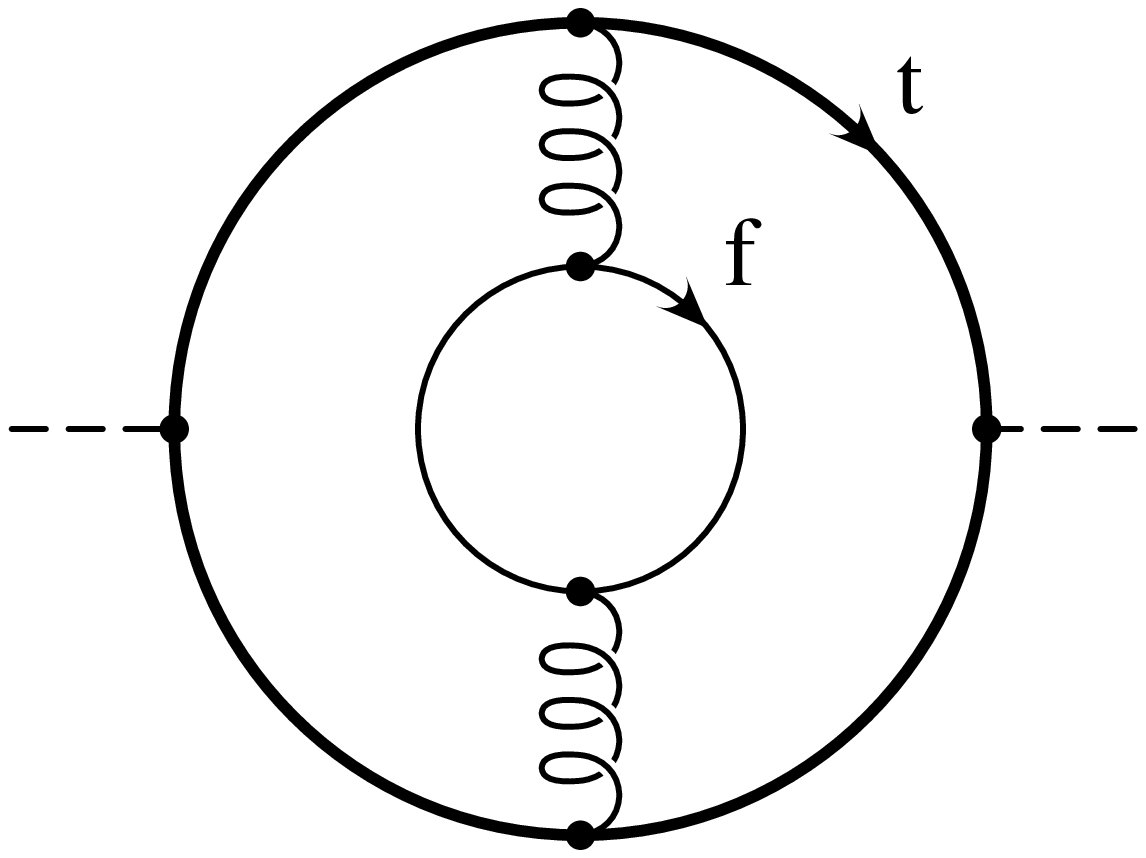}
   &
   \epsfxsize=5.0cm
   \leavevmode
   \epsffile[130 260 470 530]{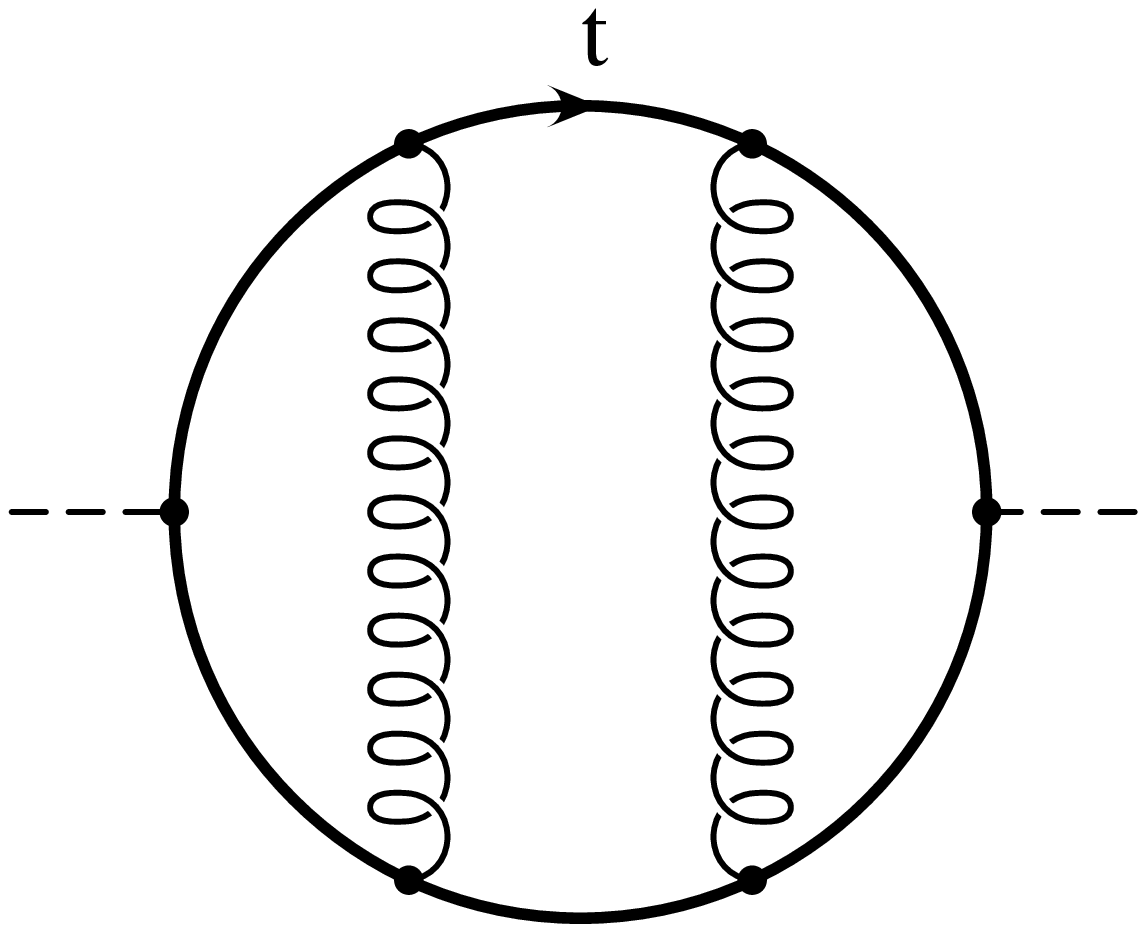}
   &
   \epsfxsize=5.0cm
   \leavevmode
   \epsffile[130 260 470 530]{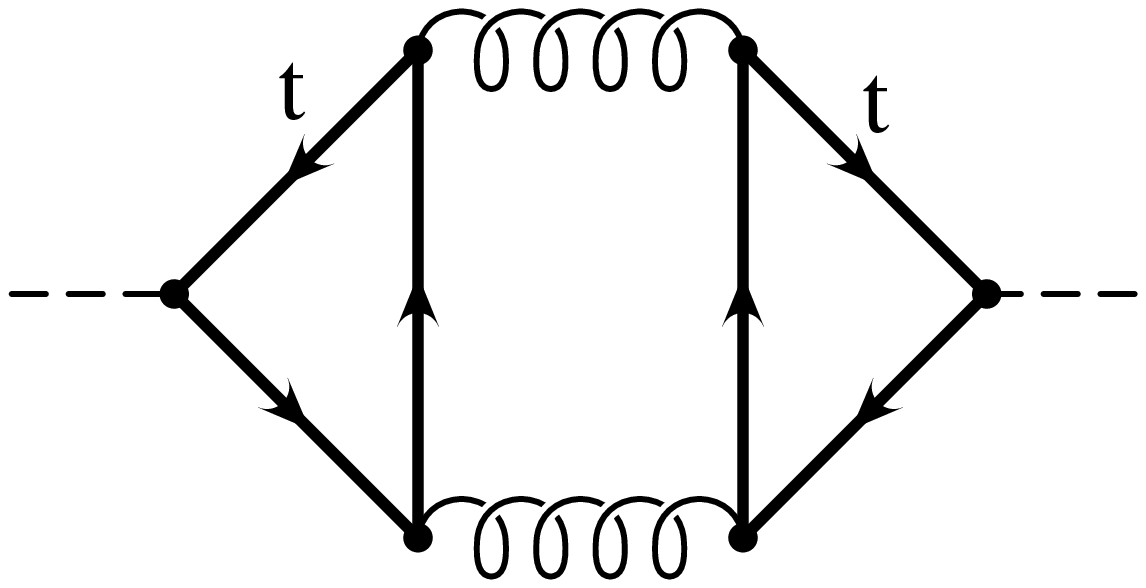}
 \end{tabular}
  \caption{\label{one}Typical Feynman diagrams pertinent to $\Pi_{HH}(q^2)$ in
  ${\cal O}(\alpha_s^2G_FM_t^2)$.
  $f$ stands for any quark.}
 \end{center}
\end{figure}

The Feynman diagrams pertinent to $\Pi_{HH}(q^2)$ in
${\cal O}(\alpha_s^2G_FM_t^2)$ come in twenty different topologies.
Typical specimen are depicted in Fig.~\ref{one}.
Using dimensional regularization, with $n=4-2\epsilon$ space-time dimensions
and a 't~Hooft mass $\mu$, and adopting from Ref.~\cite{gra} the QCD coupling
and mass counterterms in the $\overline{\mbox{MS}}$ scheme, we find
\begin{eqnarray}
\label{pihh}
\Pi_{HH}^\prime(0)&\n=\n&{3G_Fm_t^2(\mu)\over8\pi^2\sqrt2}
\left\{{2\over\epsilon}+2l-{4\over3}
+a\left(-{2\over\epsilon^2}+{5\over3\epsilon}+2l^2-{10\over3}l-{37\over18}
\right)
\right.\nonumber\\
&\n+\n&
a^2\left[{5\over2\epsilon^3}-{79\over12\epsilon^2}-{1\over3\epsilon}
\left(\zeta(3)-{311\over 36}\right)
+{5\over2}l^3-{7\over2}l^2-l\left(\zeta(3)+{1073\over72}\right)
\right.\nonumber\\
&\n-\n&\left.\left.
{16\over3}\li\left({1\over2}\right)+{11\over3}\zeta(4)+{37\over9}\zeta(3)
+{4\over3}\zeta(2)\ln^22-{2\over9}\ln^42+{17\over54}\right]\right\},
\end{eqnarray}
where $m_t(\mu)$ is the top-quark $\overline{\mbox{MS}}$ mass,
$l=\ln[\mu^2/m_t^2(\mu)]$, $\li{}$ is the quadrilogarithm, and $\zeta$ is
Riemann's zeta function.
In Eq.~(\ref{pihh}), we have omitted terms containing $\gamma_E-\ln(4\pi)$,
where $\gamma_E$ is Euler's constant.
These may be retrieved by substituting $\mu^2\to4\pi e^{-\gamma_E}\mu^2$.
We observe that, up to an overall minus sign, the ultraviolet divergences in
Eq.~(\ref{pihh}) precisely match those of the corresponding expression for
$\Pi_{WW}(0)/M_W^2$ in Ref.~\cite{avd}.
In the following, we shall employ $M_t$ instead of $m_t(\mu)$, since $M_t$
directly corresponds to the parameter which is being extracted from experiment
\cite{abe}.
The two-loop relation between $M_t$ and $m_t(M_t)$ may be found in
Ref.~\cite{gra}, and the $\mu$ evolution of $m_t(\mu)$ is determined by the
respective renormalization-group (RG) equation.

The QCD corrections to the $\ell^+\ell^-H$ Yukawa coupling originate in the
renormalizations of the Higgs-boson wave function and vacuum expectation
value.
For $M_H\ll M_t$, they may be accommodated in the $\ell^+\ell^-H$ interaction
Lagrangian by writing \cite{hff}
\begin{equation}
{\cal L}_{\ell\ell H}=-2^{1/4}G_F^{1/2}m_\ell\bar\ell\ell H(1+\delta_u),
\end{equation}
where
\begin{equation}
\delta_u=-{1\over2}\left[{\Pi_{WW}(0)\over M_W^2}+\Pi_{HH}^\prime(0)\right]
\end{equation}
is manifestly finite, gauge independent, and RG invariant.
Here, the subscript $u$ is to remind us that this term appears as a universal
building block in the radiative corrections to all production and decay
processes of the Higgs boson.
Combining Eq.~(\ref{pihh}) with the corresponding expression for
$\Pi_{WW}(0)/M_W^2$ in Ref.~\cite{avd} and eliminating $m_t(\mu)$ in favour of
$M_t$, we obtain
\begin{equation}
\label{duos}
\delta_u={7\over2}X_t\left[1-1.797\,105\,a\left(1+{7\over4}aL\right)
-16.200\,847\,a^2\right].
\end{equation}
Equation~(\ref{duos}) reproduces the well-known ${\cal O}(G_FM_t^2)$
\cite{hff,cha} and ${\cal O}(\alpha_sG_FM_t^2)$ \cite{hll} terms.
The analytic version of Eq.~(\ref{duos}) for $N_c$ arbitrary and in terms of
fundamental functions and one master integral, which may be solved numerically
with high precision \cite{avd}, will be included in Ref.~\cite{long}.

Next, we shall derive the ${\cal O}(\alpha_s^2G_FM_t^2)$ correction to the
low-$M_H$ effective $W^+W^-H$ interaction Lagrangian.
In contrast to the $\ell^+\ell^-H$ case, we are now faced with the task of
computing genuine three-point amplitudes at three loops, which, at first sight,
appears to be enormously hard.
In fact, we are not aware of any three-loop calculation of a three-point
function in the literature.
Fortunately, in the limit that we are interested in, this problem may be
reduced to one involving just three-loop two-point diagrams by means of a
low-energy theorem, whose lowest-order version has been introduced in
Refs.~\cite{ell,vai}.
Generally speaking, this theorem relates the amplitudes of two processes which
differ by the insertion of an external Higgs-boson line carrying zero
four-momentum.
It allows us to compute a loop amplitude with an external Higgs boson which is
light compared to the virtual particles by differentiating the respective
amplitude without that Higgs boson with respect to the virtual-particle masses.
In Refs.~\cite{ks1,kil}, it has been shown how the applicability of this
theorem may be extended beyond the leading order.
Proceeding along the lines of Refs.~\cite{ks2,ks3}, we obtain
\begin{equation}
\label{lwwh}
{\cal L}_{W^+W^-H}=2^{5/4}G_F^{1/2}M_W^2W_\mu^+W^{-\mu}H(1+\delta_{WWH}),
\end{equation}
with
\begin{equation}
\label{dwwhnu}
\delta_{WWH}=\delta_u+\left[1-{(m_t^0)^2\partial\over\partial(m_t^0)^2}\right]
{\Pi_{WW}(0)\over M_W^2},
\end{equation}
where $m_t^0$ is the bare top-quark mass.
In Ref.~\cite{avd}, $\Pi_{WW}(0)$ is expressed in terms of $m_t(\mu)$.
Thus, we have to undo the top-quark mass renormalization \cite{gra} before we
can apply Eq.~(\ref{dwwhnu}).
Then, after evaluating the right-hand side of Eq.~(\ref{dwwhnu}), we introduce
$M_t$ and so obtain
\begin{equation}
\delta_{WWH}=-{5\over2}X_t\left[1-2.284\,053\,a\left(1+{7\over4}aL\right)
-10.816\,384\,a^2\right].
\end{equation}
We recover the well-known ${\cal O}(G_FM_t^2)$ \cite{cha,hww}
and ${\cal O}(\alpha_sG_FM_t^2)$ \cite{ks3} terms.

The derivation of the ${\cal O}(\alpha_s^2G_FM_t^2)$ correction to the
low-$M_H$ effective $ZZH$ interaction Lagrangian proceeds in close analogy to
the $W^+W^-H$ case, and we merely list the result:
\begin{equation}
{\cal L}_{ZZH}=2^{1/4}G_F^{1/2}M_Z^2Z_\mu Z^\mu H(1+\delta_{ZZH}),
\end{equation}
where
\begin{equation}
\label{dzzhos}
\delta_{ZZH}=-{5\over2}X_t\left[1-4.684\,053\,a\left(1+{7\over4}aL\right)
-6.846\,779\,a^2\right].
\end{equation}
Equation~(\ref{dzzhos}) contains the well-known ${\cal O}(G_FM_t^2)$
\cite{cha,hzz} and ${\cal O}(\alpha_sG_FM_t^2)$ \cite{ks2} terms.

We have presented the three-loop ${\cal O}(\alpha_s^2G_FM_t^2)$ corrections to
the effective Lagrangians for the interactions of light Higgs bosons with pairs
of charged leptons, $W$ bosons, and $Z$ bosons in the SM.
As a corollary, we note that $\Gamma(H\to\ell^+\ell^-)$, $\Gamma(H\to W^+W^-)$,
and $\Gamma(H\to ZZ)$ receive the correction factors $(1+\delta_u)^2$,
$(1+\delta_{WWH})^2$, and $(1+\delta_{ZZH})^2$, respectively.
Moreover, these results may be used to refine the theoretical predictions for a
variety of four- and five-point production and decay processes of light Higgs
bosons at present and future $e^+e^-$ colliders.
This will be done in our forthcoming report \cite{long}.

Here, we would like to focus attention on an interesting theoretical point.
In fact, our analysis allows us to recognize a certain universal pattern in the
structure of the QCD perturbation series.
In addition to $\Delta\rho$, we have now three more independent observables
with quadratic $M_t$ dependence at our disposal for which the QCD expansion is
known up to next-to-leading order, namely $\delta_u$, $\delta_{WWH}$, and
$\delta_{ZZH}$.
In the on-shell scheme of electroweak and QCD renormalization, these four
electroweak parameters exhibit striking common properties.
In fact, the leading- and next-to-leading-order QCD corrections act in the
same direction and screen the ${\cal O}(G_FM_t^2)$ terms.
Furthermore, the sets of $\alpha_s/\pi$ and $(\alpha_s/\pi)^2$ coefficients
each lie in the same ball park.
{}For the choice $\mu=M_t$, the coefficients of $\alpha_s/\pi$ range between
$-1.797$ and $-4.684$, and those of $(\alpha_s/\pi)^2$ between $-6.847$ and
$-16.201$.
We would like to point out that the corresponding coefficients of the ratio
$\mu_t^2/M_t^2$, where $\mu_t=m_t(\mu_t)$, are $-2.667$ and $-11.140$
\cite{long}, {\it i.e.}, they lie right in the centres of these ranges.
Therefore, it suggests itself that the use of $M_t$ is the origin of these
similarities.
In fact, if we express the QCD expansions in terms of $\mu_t$ rather than $M_t$
and choose $\mu=\mu_t$, then the coefficients of $\alpha_s/\pi$ and
$(\alpha_s/\pi)^2$ nicely group themselves around zero;
they range from $-2.017$ to 0.870 and from $-3.970$ to $1.344$, respectively
\cite{long}.
This indicates that the perturbation expansions converge more rapidly if we
renormalize the top-quark mass according to the $\overline{\mbox{MS}}$ scheme.
Without going into details, we would like to mention that the study of
renormalons \cite{ren} offers a possible theoretical explanation of this
observation.
Since the on-shell and $\overline{\mbox{MS}}$ results coincide in lowest
order, this does, of course, not imply that the QCD corrections are any
smaller in the $\overline{\mbox{MS}}$ scheme.
It just means that, as a rule, the ${\cal O}(G_FM_t^2)$ terms with $M_t$
replaced by the two-loop expression for $\mu_t$ \cite{long} are likely to
provide fair approximations for the full three-loop results.
In all the cases considered here, the QCD corrections now appear to be well
under control.

\bigskip

We would like to thank Bill Bardeen, Kostja Chetyrkin, and Michael Spira for
very useful discussions.
One of us (BAK) is indebted to the FNAL Theory Group for inviting him as a
Guest Scientist.
He is also grateful to the Phenomenology Department of the University of
Wisconsin at Madison for the great hospitality extended to him during a
visit when a major part of his work on this project was carried out.

\newpage

\begin{thebibliography}{99}
%
\bibitem{abe} CDF Collaboration, F. Abe {\it et al.},
Phys.\ Rev.\ Lett.\ {\bf74}, 2626 (1995);
D0 Collaboration, S. Abachi {\it et al.},
Phys.\ Rev.\ Lett.\ {\bf74}, 2632 (1995).
%
\bibitem{bjo} J.D. Bjorken, in {\it Weak Interactions at High Energy and the
Production of New Particles: Proceedings of Summer Institute on Particle
Physics}, August~2--13, 1976, edited by M.C. Zipf, SLAC Report No.~198 (1976)
p.~1.
%
\bibitem{jan} P. Janot, lecture delivered at {\it First General Meeting of the
LEP2 Workshop}, CERN, Geneva, Switzerland, 2--3 February 1995.
%
\bibitem{ell} J. Ellis, M.K. Gaillard, and D.V. Nanopoulos,
Nucl.\ Phys.\ {\bf B106}, 292 (1976).
%
\bibitem{pr} B.A. Kniehl, Phys.\ Rep.\ {\bf240}, 211 (1994).
%
\bibitem{hll} B.A. Kniehl and A. Sirlin, Phys.\ Lett.\ B {\bf318}, 367 (1993);
B.A. Kniehl, Phys.\ Rev.\ D {\bf50}, 3314 (1994);
A. Djouadi and P. Gambino, Phys.\ Rev.\ D {\bf51}, 218 (1995).
%
\bibitem{ks1} B.A. Kniehl and M. Spira, Nucl.\ Phys.\ {\bf B432}, 39 (1994).
%
\bibitem{kwi} A. Kwiatkowski and M. Steinhauser,
Phys.\ Lett.\ B {\bf338}, 66 (1994); {\bf342}, 455(E) (1995).
%
\bibitem{ks2} B.A. Kniehl and M. Spira, Nucl.\ Phys.\ {\bf B443}, 37 (1995).
%
\bibitem{ks3} B.A. Kniehl and M. Spira, Report Nos.\ DESY~95--041,
FERMILAB--PUB--95/081--T, MPI/PhT/95--21, hep--ph/9505225,
Z. Phys.\ C (in press).
%
\bibitem{gam} A. Djouadi and P. Gambino,
Phys.\ Rev.\ Lett.\ {\bf73}, 2528 (1994).
%
\bibitem{avd} L. Avdeev, J. Fleischer, S. Mikhailov, and O. Tarasov,
Phys.\ Lett.\ B {\bf336}, 560 (1994); {\bf349}, 597(E) (1995);
K.G. Chetyrkin, J.H. K\"uhn, and M. Steinhauser,
Phys.\ Lett.\ B {\bf351}, 331 (1995);
in {\it Proceedings of the Ringberg Workshop on Perspectives of
Electroweak Interactions in $e^+e^-$ Collisions}, Tegernsee, Germany,
February 5--8, 1995, edited by B.A. Kniehl
(World Scientific, Singapore, 1995).
%
\bibitem{djo} A. Djouadi and C. Verzegnassi,
Phys.\ Lett.\ B {\bf195}, 265 (1987);
A. Djouadi, Nuovo Cim.\ {\bf100A}, 357 (1988);
B.A. Kniehl, Nucl.\ Phys.\ {\bf B347}, 86 (1990).
%
\bibitem{sir} A. Sirlin, Phys.\ Rev.\ D {\bf22}, 971 (1980);
K-I. Aoki, Z. Hioki, R. Kawabe, M. Konuma, and T. Muta,
Prog.\ Theor.\ Phys.\ Suppl.\ {\bf73}, 1 (1982);
M. B\"ohm, H. Spiesberger, and W. Hollik,
Fortschr.\ Phys.\ {\bf34}, 687 (1986).
%
\bibitem{msb} W.A. Bardeen, A.J. Buras, D.W. Duke, and T. Muta,
Phys.\ Rev.\ D {\bf18}, 3998 (1978).
%
\bibitem{long} B.A. Kniehl and M. Steinhauser, in preparation.
%
\bibitem{gra} N. Gray, D.J. Broadhurst, W. Grafe, and K. Schilcher,
Z. Phys.\ C {\bf48}, 673 (1990).
%
\bibitem{hff} B.A. Kniehl, Nucl.\ Phys.\ {\bf B376}, 3 (1992).
%
\bibitem{cha} M.S. Chanowitz, M.A. Furman, and I. Hinchliffe,
Phys.\ Lett.\ {\bf78B}, 285 (1978); Nucl.\ Phys.\ {\bf B153}, 402 (1979);
Z. Hioki, Phys.\ Lett.\ B {\bf224}, 417 (1989); {\bf228}, 560(E) (1989).
%
\bibitem{vai} A.I. Va\u\i nshte\u\i n, M.B. Voloshin, V.I. Zakharov, and
M.A. Shifman,
Yad.\ Fiz.\ {\bf30}, 1368 (1979) [Sov.\ J. Nucl.\ Phys.\ {\bf30}, 711 (1979)];
A.I. Va\u\i nshte\u\i n, V.I. Zakharov, and M.A. Shifman,
Usp.\ Fiz.\ Nauk {\bf131}, 537 (1980) [Sov.\ Phys.\ Usp.\ {\bf23}, 429 (1980)];
L.B. Okun, {\it Leptons and Quarks}, (North-Holland, Amsterdam, 1982);
M.B. Voloshin, Yad.\ Fiz.\ {\bf44}, 738 (1986)
[Sov.\ J. Nucl.\ Phys.\ {\bf44}, 478 (1986)];
M.A. Shifman, Usp.\ Fiz.\ Nauk {\bf157}, 561 (1989)
[Sov.\ Phys.\ Usp.\ {\bf32}, 289 (1989)];
J.F. Gunion, H.E. Haber, G. Kane, and S. Dawson,
{\it The Higgs Hunter's Guide}, (Addison-Wesley, Redwood City, 1990).
%
\bibitem{kil} W. Kilian, Report Nos.\ DESY~95--075 and hep--ph/9505309
(May 1995).
%
\bibitem{hww} B.A. Kniehl, Nucl.\ Phys.\ {\bf B357}, 439 (1991).
%
\bibitem{hzz} B.A. Kniehl, Nucl.\ Phys.\ {\bf B352}, 1 (1991).
%
\bibitem{ren} M. Beneke and V.M. Braun, Phys.\ Lett.\ B {\bf348}, 513 (1995);
M. Neubert, Phys.\ Rev.\ D {\bf51}, 5924 (1995);
Report Nos.\ CERN--TH.7524/94 and hep--ph/9502264 (revised April 1995);
P. Ball, M. Beneke, and V.M. Braun, Report Nos.\ CERN--TH/95--26,
UM--TH--95--3, and hep--ph/9502300 (February 1995);
K. Philippides and A. Sirlin, Report Nos.\ NYU--TH--95/03/01 and
hep--ph/9503434 (March 1995);
P. Gambino and A. Sirlin, Report Nos.\ NYU--TH--95/05/02 and
hep--ph/9505426 (May 1995).
%
\end{thebibliography}
\end{document}